\pdfoutput=1
\documentclass[12pt,prx,aps]{revtex4-2}

\usepackage{amsmath}
\usepackage{amssymb}
\usepackage{mathrsfs}
\usepackage{graphicx}
\usepackage{bm}
\usepackage{comment}

\newcommand\+{\dagger}
\newcommand\<{\langle}
\renewcommand\>{\rangle}
\renewcommand\d{\partial}
\newcommand\x{\mathbf{x}}
\newcommand\y{\mathbf{y}}
\newcommand\p{\mathbf{p}}

\begin{document}
\title{Chiral Metric Hydrodynamics, Kelvin Circulation Theorem, and the Fractional Quantum Hall Effect}

\date{July 2019}

\author{Dam Thanh Son}
%\email{dtson@uchicago.edu}
%\affiliation{Department of Physics, University of Chicago, Chicago, Illinois 60637, USA}
\affiliation{Kadanoff Center for Theoretical Physics, University of Chicago, Chicago, Illinois 60637, USA}

\begin{abstract}
  By extending the Poisson algebra of ideal hydrodynamics to include a
  two-index tensor field, we construct a new (2+1)-dimensional
  hydrodynamic theory that we call ``chiral metric hydrodynamics.''
  The theory breaks spatial parity and contains a degree of freedom
  which can be interpreted as a dynamical metric, and describes a medium which
  behaves like a solid at high frequency and a fluid with odd
  viscosity at low frequency.  We derive a version of the Kelvin
  circulation theorem for the new hydrodynamics, in which the
  vorticity is replaced by a linear combination of the vorticity and
  the dynamical Gaussian curvature density.  We argue that the chiral metric
  hydrodynamics, coupled to a dynamical gauge field, correctly
  describes the long-wavelength
  dynamics of quantum Hall Jain states with filling factors $\nu=N/(2N+1)$
  and $\nu=(N+1)/(2N+1)$ at large $N$.  The Kelvin circulation theorem
  implies a relationship between the electron density and the dynamical Gaussian
  curvature density.  We present
  a purely algebraic derivation of the low-momentum asymptotics of
  the static structure factor of the Jain states.
\end{abstract}
\maketitle

\section{Introduction}

As a general framework, hydrodynamics~\cite{LL6} finds its application
in a wide range of physical contexts, including classical systems like
classical fluids, gases, or liquid crystals and systems where quantum
mechanics plays an important role like superfluids or quantum
electronic fluids.  Recently, attention has been drawn to systems
where the hydrodynamic description manifests features resembling or
connected to the topological properties of quantum systems.  One
example is fluids with odd (or Hall)
viscosity~\cite{Avron:1995fg,Avron:1997}, which can exist in two
spatial dimensions if spatial reflection symmetry is absent.  In the
most well-known example of such a fluid---the electron fluid in the
quantum Hall effect---the odd viscosity is proportional to the shift, a
topological property of the quantum Hall states~\cite{Read:2008rn}.
Another example is fluids of chiral fermions in three dimensions,
where the presence of triangle anomalies in the microscopic theory
give rise to modifications of the hydrodynamic equations of the finite
temperature system, with important physical
consequences~\cite{Son:2009tf}.

In this paper, we describe a theory that we term ``chiral metric
hydrodynamics''---an extension of the hydrodynamic theory which
includes a tensorial degree of freedom, which, for certain purposes, can be
interpreted as a dynamical metric.  One motivation for considering such a theory
is to understand fractional quantum Hall fluids, where the lowest
gapped mode is a spin-two mode, the so-called ``magnetoroton''
\cite{Girvin:1986zz}.
%Near a nematic phase transition the
%magnetoroton is parametrically lighter than all other modes.
In Ref.~\cite{Gromov:2017qeb} it was proposed that the theory
governing the dynamics of the magnetoroton at long wavelength is a
theory of a dynamical metric, as suggested originally by
Haldane~\cite{Haldane:2009ke}.  One aim of this work is to relate the
ideas of Refs.~\cite{Haldane:2009ke,Gromov:2017qeb} with the seemingly
unrelated picture of a quantum Hall fluid as a fluid of composite
fermions obeying a particular form of
hydrodynamics~\cite{Stone:1990cd}.  (A phenomenological model that
exhibits some features of the theory presented in this paper has been
considered in Ref.~\cite{Tokatly:2005}.)

%We show that the ``metric hydrdodynamics'' possesses many interesting
%properties and also can also be applied to a number of physical
%situations.

We will first provide a purely mathematical construction of the chiral
metric hydrodynamics by enlarging the Poisson algebra of
hydrodynamics with the inclusion of a two-index tensor.  We will show
that, while in general dimensions the resulting theory is simply the
theory of elasticity, in two spatial dimensions there is a possibility
of a parity-violating Poisson algebra, leading to a theory of a chiral
viscoelastic medium which behaves like a solid at high frequencies,
but like a fluid with odd viscosity at low frequencies.  The
hydrodynamic theory describing this medium has interesting features,
in particular there is a version of the Kelvin circulation theorem
involving a particular linear combination of the vorticity and the
density of the Gaussian curvature of the dynamical metric.

Coupling the fluid to a dynamical gauge field, the theory, as we will
show, can be used to describe the low-energy dynamics of the fluid of
composite fermions in the fractional quantum Hall effect, in
particular in the Jain states with filling factors $\nu=N/(2N+1)$ and
$\nu=(N+1)/(2N+1)$ at large $N$.  By using the Kelvin circulation
theorem and the fact that the composite fermion carries an electric
dipole moment, we will establish a relationship between the electron
density and the dynamical Gaussian curvature density,
and use this connection to compute the statics structure factor of
quantum Hall state.

%\section{Nematric hydrodynamics}

\section{The Poisson algebra}

\subsection{The Poisson algebra of classical hydrodynamics}

We recall that the equations of compressible fluid dynamics can be
recasted in the form of Hamilton's equations of motion of a dynamical
system with a Hamiltonian and a set of Poisson brackets.  In a normal
fluid, the hydrodynamic degrees of freedom are the particle-number
density $n(\x)$ and the momentum density $\bm{\pi}(\x)$.  The Poisson
brackets between these degrees of freedom are~\cite{Landau41}
\begin{subequations}\label{hydro-PB}
\begin{align}
  \{ \pi_i(\x), \, n(\y) \} &= n(\x) \d_i \delta(\x-\y), \\
  \{ \pi_i(\x), \, \pi_j(\y) \}
  &= [\pi_j(\x)\d_i + \pi_i(\y)\d_j]\delta(\x-\y).
%  - \epsilon_{ij} b(\x) n(\x)\delta(\x-\y)
\end{align}
\end{subequations}
%These Poisson brackets can be derived from the expressions for the
%density and momentum density of a system of $N$ particles with
%coordinates $\x_a$ and momenta $\p_a$, $a=1,\ldots, N$~\cite{Landau41}
%\begin{equation}
%  n(\x) = \sum_{a=1}^N \delta(\x-\x_a), \qquad
%  \pi_i(\x) = \sum_{a=1}^N \p_a \delta(\x-\x_a)
%\end{equation}
%and the canonical Poisson brackets~\cite{LL1}
%\begin{equation}
%  \{ p_{ai},\, x_b^j \} = \delta_{ab} \delta_i^j
%\end{equation}
The Poisson algebra, together with a Hamiltonian in the form of a
functional of the hydrodynamic degrees of freedom
\begin{equation}
  H = H [n(\x), \pi_i(\x)],
\end{equation}
completely determines the time evolution of the hydrodynamic fields:
$\dot O=\{H,O\}$, with $O$ being $n$ or $\pi_i$.  Denoting the
variational derivatives of $H$ with respect to $n(\x)$ and $\pi_i(\x)$
by $\mu(\x)$ and $v^i(\x)$, respectively,
\begin{equation}
  \delta H = \int\!d\x \left[\mu(\x) \delta n(\x)  + v^i(\x) \delta \pi_i(\x) \right],
\end{equation}
the equations of motion can be written as
\begin{subequations}\label{hydro_eq}
\begin{align}
  & \dot n + \d_i (nv^i) = 0 , \\
  & \dot \pi_i + n\d_i\mu + \pi_j\d_i v^j + \d_j(v^j\pi_i) = 0.
\end{align}
\end{subequations}
In the case of a fluid with Galilean invariance, where the Hamilton
is the sum of the kinetic and potential energies:
\begin{equation}
  H = \int\!d\x\, \left[ \frac{\bm{\pi}^2}{2m n} + \epsilon(n) \right],
\end{equation}
with $m$ being the mass of the fluid particles,
Eqs.~(\ref{hydro_eq}) coincide with the Euler equations for an
ideal fluid.

\subsection{Poisson algebra and spatial diffeomorphism. Introducing the metric}

We will now introduce into the
formalism a new hydrodynamic variable---a two-index tensor $G_{ij}$,
which will also be called the ``metric.''  The introduction of a
tensor into hydrodynamics makes our theory similar to hydrodynamics of
nematic liquid crystals, for which the
%although the
%applications that we will discuss later are all of quantum nature.
%---we will comment on
%how our theory differs from the existing nematic hydrodynamic theories
%later.  For now,
Poisson bracket formalism has been applied~\cite{StarkLubensky}.  Here
we first abstract ourselves from physical applications and construct
our theory from purely mathematical consideration.

First, we note that the momentum density $\bm{\pi}(\x)$ can be
regarded as the operator generating infinitesimal diffeomorphism
transformation~\cite{DzyaloshinskiiVolovick}.  More concretely, if one
defines for each vector field $\xi^k(\x)$ the quantity
\begin{equation}
  \hat\xi = \int\!d\y\, \xi^k(\y)\pi_k(\y),
\end{equation}
then the Poisson brackets of $\hat\xi$ with $n$ and $\pi_i$,
\begin{subequations}\label{PB-xi-npi}
\begin{align}
  \{ \hat\xi, \, n(\x) \} &= -\xi^k \d_k n - n\d_k\xi^k
%     = -\pounds_\xi n - n\d_k\xi^k
     , \\
  \{\hat\xi, \, \pi_i(\x) \} &=
     -\xi^k\d_k\pi_i - \pi_k\d_i\xi^k - \xi_i \d_k \xi^k
%     = -\pounds_\xi \pi_i - \pi_i \d_k\xi^k
     ,
\end{align}
\end{subequations}
reproduce exactly the shift of a scalar density and a covariant
vector density (both carrying weight $w=1$) under an infinitesimal coordinate transformation
$x^k\to x^k+\xi^k$, and
can be written compactly through the Lie derivative\footnote{Recall that the Lie derivative a tensor density $T_{i_1\ldots i_m}^{j_1\ldots j_n}$ of weight $w$ with respect to $\xi$ is
\begin{align*}
  (\pounds_\xi T)_{i_1\ldots i_m}^{j_1\ldots j_n} =
  \xi^k \d_k T_{i_1\ldots i_m}^{j_1\ldots j_n}
  & + T_{k\ldots i_m}^{j_1\ldots j_n} \d_{i_1}\xi^k
%  + T_{i_1k\ldots i_m}^{j_1j_2\ldots j_n} \d_{i_2}\xi^k
  + \cdots
  + T_{i_1\ldots k}^{j_1\ldots j_n} \d_{i_m}\xi^k\\
  & - T_{i_1\ldots i_m}^{k\ldots j_n} \d_k \xi^{j_1}
%  - T_{i_1i_2\ldots i_m}^{j_1k\ldots j_n} \d_k \xi^{j_2}
  - \cdots
  - T_{i_1\ldots i_m}^{j_1\ldots k} \d_k \xi^{j_n}
  + w T_{i_1\ldots i_m}^{j_1\ldots j_n}\d_k \xi^k
\end{align*}
}:
\begin{equation}
  \{ \hat\xi, \, O(\x) \} = \pounds_\xi O, \qquad O = n, \pi_i
\end{equation}

In other words, the Poisson algebra formed by
$\pi_i(\x)$ is isomorphic to the algebra of spatial diffeomorphism,
and the Poisson bracket of $\pi_i$ with any quantity is determined by
the law of transformation of the latter under diffeomorphism.  For
example, the momentum per particle $u_i=\pi_i/n$ is a vector, and this is
manifested in its Poisson bracket with the momentum density:
\begin{equation}
  \{ \hat\xi, \, u_i (\x) \} = -\xi^k \d_k u_i - u_k \d_i\xi^k  =
  -\pounds_\xi u_i \,.
\end{equation}

We are now ready to introduce a new degree of freedom into the theory:
the metric $G_{ij}$.  It is symmetric, $G_{ij}=G_{ji}$, and we chose it transform as a
covariant tensor, which means
%We demand that the commutator of $\hat v$ with $G_{ij}$ reflects the
%tensor nature of the latter:
\begin{equation}
  \{\hat\xi,\, G_{ij}(\x)\} =  - \pounds_\xi G_{ij} = 
  -\xi^k\d_k G_{ij} - G_{kj}\d_i\xi^k - G_{ik}\d_j\xi^k .
\end{equation}
This condition fixes the Poisson bracket between $\pi_i$ and $G_{ij}$,
\begin{equation}
  \{ G_{ij}(\x), \, \pi_k(\y) \} = \left[ G_{ik}(\x)\d_j + G_{jk}(\x)\d_i 
    + \d_k G_{ij}(\x) \right]\delta(\x-\y) .
\end{equation}
Note that this is quite different from the corresponding Poisson
bracket in the theory of nematic liquid crystals~\cite{StarkLubensky}.

As in general relativity,
we will denote the matrix inverse of $G_{ij}$ as $G^{ij}$ and $\det
G_{ij}$ as $G$.  Note that $G$ transforms like a scalar density, i.e.,
in the same way as $n$.  Thus it is consistent to impose the
constraint $\sqrt{G}=n$.  This means that beside $n$ and $\bm\pi$, the
new dynamical degrees of freedom form a unimodular matrix
$G_{ij}/n^{2/d}$, where $d$ is the number of spatial dimensions.

To complete the Poisson algebra, we need to define the Poisson
brackets between $G_{ij}$ and $n$, and between the components of
$G_{ij}$.   For the Poisson bracket between $G$ and $n$ the simplest
choice is to declare it to vanish:
\begin{equation}
  \{ G_{ij}(\x), \, n(\y) \} = 0 .
\end{equation}
As to the Poisson bracket between components of $G_{ij}$, there are
two different natural choices that lead to dramatically different
hydrodynamic theories.  The simpler choice, which works in any number
of spatial dimensions, is to take the Poisson brackets between the
metric components to be zero,
\begin{equation}
  \{ G_{ij}(\x),\, G_{kl}(\y) \} = 0 .
\end{equation}
However, in this case our hydrodynamic theory is simply another
formulation of the theory of elasticity.  This can be seen as follows.
Assume the most general Hamiltonian $H = H[n(\x),\pi_i(\x),
  G_{ij}(\x)]$ and introducing again its functional derivatives,
\begin{equation}\label{deltaH}
  \delta H = \int\!d\x\, \left[ \mu(\x) \delta n(\x) + v^i(\x)\delta\pi_i(\x)
    + \frac1{2} \sigma^{ij}(\x) \delta G_{ij}(\x) \right],
\end{equation}
the equations of motion have the form
\begin{subequations}\label{hydro_eqs_parity}
\begin{align}
  \d_t n &= - \d_i(nv^i) , \\
  \d_t \pi_i &= -n\d_i\mu -\pi_j \d_i v^j -\d_j(v^j\pi_i)
  - \d_j (\sigma^{jk} G_{ki})
  + \frac12 \sigma^{jk}\d_i G_{jk} \,, \\
  \d_t G_{ij} &= -v^k \d_k G_{ij} - G_{ik}\d_j v^k - G_{jk}\d_i v^k .
%  + \frac1s (\epsilon_{ik} G_{jk} + \epsilon_{jk} G_{il}) \tau^{kl}
  \label{dtG_nos}
\end{align}
\end{subequations}
%The equation of momentum conservation can be written as
%\begin{equation}
%  \d_t\pi_i + \d_j (v^j\pi_i + \tau^{jk}G_{ki} + p\delta^j_i) = 0
%\end{equation}
In particular, Eq.\ (\ref{dtG_nos}) means that the metric tensor is
frozen into the fluid, and is given simply by the shape of an
initially spherical droplet advected by the flow.  The fact that the
energy depends on the shape of this droplet means that we are dealing
not with fluid, but with an elastic solid, where $\sigma^{ij}$ is
basically the elastic stress.  This finding can be confirmed, for
example, by finding the normal modes of the linearized hydrodynamic
equations: one finds, in addition to the longitudinal sound, the
transverse sound.

\section{Chiral Poisson algebra and chiral metric hydrodynamics}

\subsection{Construction of the chiral metric hydrodynamic theory}

In two spatial dimensions, if one does not impose spatial parity, then
there is another choice for the Poisson bracket between $G_{ij}$:
\begin{equation}\label{GG-PB}
  \{ G_{ij}(\x), \, G_{kl}(\y)\} = - \frac1s \left( \epsilon_{ik}G_{jl}
  + \epsilon_{il}G_{jk} + \epsilon_{jk} G_{il}
  + \epsilon_{jl} G_{ik} \right) \delta(\x-\y),
\end{equation}
where $s$ is a pure number, and $\epsilon_{ij}=-\epsilon_{ji}$,
$\epsilon_{12}=1$.
One needs to verify that the Jacobi identity is satisfied.
Excluding the delta-function on the right-hand
side, Eq.~(\ref{GG-PB}) has the same form as the commutators of the
$\mathfrak{sl}(2,\mathbb{R})$ algebra.
%The right-hand side
%of Eq.~(\ref{GG-PB}) is the unique expression that can be written in
%terms of the $G_{ij}$ and the 
That means that the Jacobi identity $\{\{G,G\},G\}+\cdots=0$ is
satisfied.  Checking the Jacobi identity $\{\pi,\{G,G\}\}+\cdots=0$ is
equivalent to verifying that the two sides of Eq.~(\ref{GG-PB})
transform in the same way under general coordinate transformations.
This can be established by noting that $\delta(\x-\y)$ is a scalar
density of weight $-1$ and $\epsilon_{ij}$ a tensor density of weight
$+1$.  Other cases can be checked trivially.

\begin{comment}

it is convenient to
think about $G_{ij}$ as a metric and introduce a vielbein $e^a_i$,
$a=1,2$, so that
\begin{equation}
  G_{ij} = e^a_i e^a_j
\end{equation}
The Poisson brackets involving $G$ are the consequences of the following
Poisson brackets involving the vielbein,
\begin{align}
  \{\pi_i(\x), \, e^a_j(\y) \} &= e^a_j(\x)\d_i\delta(\x-\y) \\
  \{ e^a_i(\x), \, e^b_j(\y) \} &= -\frac1s \delta^{ab}\epsilon_{ij}
    \delta(\x-\y)
\end{align}
The Jacobi identity can be now be checked with little difficulty.  In particular, the Poisson bracket $\{pi_i,\, e^a_j\}$ means that $e^a_i$ transforms like
a covariant vector, 

\end{comment}

Given the Hamiltonian $H=H[n(\x),\pi_i(\x),G_{ij}(\x)]$, the
hydrodynamic equations can be now written down using the conjugate
variables defined in Eq.~(\ref{deltaH}):
\begin{subequations}\label{hydro_eqs_noparity}
\begin{align}
  \d_t n &= - \d_i(nv^i),  \label{dtn-noparity} \\
  \d_t \pi_i &= -n\d_i\mu -\pi_j \d_i v^j -\d_j(v^j\pi_i)
  - \d_j(\sigma^{jk}G_{ki})
  + \frac12 \sigma^{jk}\d_i G_{jk}\,, \label{dtpi-noparity} \\
  \d_t G_{ij} &= -v^k \d_k G_{ij} - G_{ik}\d_j v^k - G_{jk}\d_i v^k
  + \frac1s (\epsilon_{ik} G_{jl} + \epsilon_{jk} G_{il}) \sigma^{kl} .
  \label{dtG-noparity}
\end{align}
\end{subequations}
Compared to Eqs.~(\ref{hydro_eqs_parity}), the equation for time
evolution of the metric contains a parity-odd term.

If the metric tensor $G_{ij}$ is almost isotropic, like 
in most applications that we will consider, then one can expand
\begin{equation}
  G_{ij} = n(\delta_{ij} + Q_{ij}), \qquad Q_{ij}\ll1,  ~ Q_{ii}=0 .
\end{equation}
In this limit, one can use the ``reduced'' Poisson brackets
\begin{align}
  \{ Q_{ij}(\x),\, \pi_k(\y)\} &= (\delta_{ik}\d_j+\delta_{jk}\d_i-
  \delta_{ij}\d_k)\delta(\x-\y), \\
  \{ Q_{ij}(\x),\, Q_{kl}(\y) \} &= -\frac1{sn} (
  \epsilon_{ik}\delta_{jl} + \epsilon_{il}\delta_{jk}
  + \epsilon_{jk}\delta_{il} + \epsilon_{jl}\delta_{ik} )
  \delta(\x-\y),
\end{align}
to derive the hydrodynamic equations
\begin{align}
  & \d_t n + \d_i (n v^i) = 0,\\
  & \d_t\pi_i + \d_j(v^j\pi_i) + n\d_i\mu + \pi_j\d_i v^j + \d_j \tau^{ji} = 0, \label{NStau}\\
  & \d_t Q_{ij}
%  + v^k \d_k Q_{ij} + Q_{ik}\d_j v_k + Q_{jk}\d_i v_k - Q_{ij}\d_k v_k
  -\frac1{sn}(\epsilon_{ik}\tau_{kj}+\epsilon_{jk}\tau_{ki})
  + V_{ij} = 0, \label{dtQeV}
\end{align}
where $\tau^{ij}=n\sigma^{ij}$ and
\begin{equation}
  V_{ij} = \d_i v_j + \d_j v_i - \delta_{ij}\bm{\nabla}\cdot\mathbf{v} .
\end{equation}
Note that in the equation for momentum conservation $\tau^{ij}$ plays the
role of an anisotropic stress.

\subsection{The high- and low-frequency regimes}

Ignoring for a moment the $V_{ij}$ term, if one introduces the
``Lam\'e coefficient'' $\tilde\mu$ so that $\tau^{ij}=\tilde\mu
Q_{ij}$, the equation for $Q_{ij}$
\begin{equation}
  \d_t Q_{ij} = \frac{\tilde\mu}{ns} (\epsilon_{ik} Q_{kj}
    + \epsilon_{jk} Q_{ki}),
\end{equation}
means that the metric tensor rotates in the space $(Q_{xx},Q_{xy})$
with angular velocity
\begin{equation}\label{omegaQ}
  \omega_Q = \frac{2\tilde\mu}{ns} \,.
\end{equation}
If the upper range of validity of our hydrodynamic theory
is much larger than $\omega_Q$, then one can talk about the
high-frequency ($\omega\gg\omega_Q$) and low-frequency
($\omega\ll\omega_Q$) regimes.  In the former regime Eq.~(\ref{dtQeV})
becomes $\d_t Q_{ij}+V_{ij}=0$, which can be solved in terms of the
displacement $u_i$, defined so that $v_i=\dot u_i$:
\begin{equation}
  Q_{ij} = -(\d_i u_j + \d_j u_i - \delta_{ij}\d_k u_k),
\end{equation}
and Eq.~(\ref{NStau}) becomes the equation of motion of an elastic
medium with $\tilde\mu$ being a Young modulus.

In the low-frequency regime $\omega\ll\omega_Q$, Eq.~(\ref{dtQeV})
reduces to the balance of the last two terms:
\begin{equation}
   -\frac1{sn}(\epsilon_{ik}\tau_{kj}+\epsilon_{jk}\tau_{ki})
  + V_{ij} = 0,
\end{equation}
whose solution is
\begin{equation}
  \tau^{ij} = - \frac{sn}4 \left(\epsilon_{ik}V_{kj} + \epsilon_{jk}V_{ik}\right).
\end{equation}
But $\tau^{ij}$ appears as a contribution to the stress tensor in
momentum conservation~(\ref{NStau}).  This means the fluid has a
nonzero odd (or Hall) viscosity equal to
\begin{equation}\label{etaH}
  \eta^H = \frac{sn}2 \,.
\end{equation}
Following Ref.~\cite{Read:2008rn}, $s$ can then be identified with the
average orbital spin per particle.

\begin{comment}
\subsubsection{Dirac brackets }

At low frequencies one can ``integrate out'' the massive mode $Q_{ij}$
to derive hydrodynamics with odd viscosity.  In the Hamiltonian
formalism such process of integrating out the high-energy degrees of
freedom has the effect of replacing the Poisson brackets of low-energy
degrees of freedom by the Dirac brackets.  Computing the Dirac
brackets one finds
\begin{equation}
  \{ \pi_i(\x), \, \pi_j(\y)\} = \cdots
\end{equation}
This are exactly the Poisson brackets 

Let us now find the eigenmodes in the flow.  We assume that for small
perturbations $\delta n=n-n_0$, $\pi_i$,
\begin{equation}
  H = \int\!d\x\, \left(\frac1{2mn_0}\bm{\pi}^2 + \frac\chi2 \delta n^2 +
    \frac\mu4 Q_{ij} Q_{ij}\right) 
\end{equation}
where $m$, $\chi$, and $\mu$ can be called the mass, compressibility
and Young modulus.
\end{comment}

\subsection{The Kelvin circulation theorem in chiral metric hydrodynamics}

In ordinary ideal hydrodynamics, there exist an infinite number of
conserved quantities
\begin{equation}\label{Casimirs}
  I_F = \int\!d\x\, n(\x) F\left(\frac{\omega(\x)}{n(\x)}\right),
\end{equation}
where
\begin{equation}\label{omega-def}
  \omega = \epsilon^{ij} \d_i u_j
  \equiv \epsilon^{ij} \d_i \left( \frac{\pi_j} n\right) .
\end{equation}
will be called ``vorticity,''\footnote{In a Galilean-invariant fluid,
  our definition of the vorticity differs from the usual definition by
  a factor of $m$, the mass of the fluid particle.  We emphasize,
  however, that the Kelvin circulation theorem exists without Galilean
  invariance.} and $F$ is an arbitrary function of its argument.  The
conservation of these charges does not depend on the Hamiltonian:
$I_F$ has zero Poisson brackets with all hydrodynamics
variables, i.e., $n$ and $\pi_i$.  In other words, $I_F$ are Casimirs
of the Poisson algebra.  This can established from the Poisson
brackets
\begin{align}\label{omega-PB}
  \{ n(\x), \, \omega(\y) \} &= 0,\\
  \{ \pi_i(\x), \, \omega(\y) \} &= \omega(\x) \d_i \delta(\x-\y),
  \label{piomega-PB}
\end{align}
which follow from Eq.~(\ref{hydro-PB}).  The second Poisson bracket~(\ref{piomega-PB})
is equivalent to the statement that $\omega(\x)$ transform like a
scalar density, which can already be seen from Eq.~(\ref{omega-def}).
The time evolution of $\omega$ has the form of a conservation law:
\begin{equation}
\d_t\omega+\d_i(\omega v^i)=0.
\end{equation}
from which the conservation of (\ref{Casimirs}) also follows.

In metric hydrodynamics $I_F$ are no longer Casimirs: the Poisson
bracket of the vorticity with the metric $G_{ij}$ is nonzero.  It
turns out, however, that within the chiral metric hydrodynamic theory
there exists a modified version of the Kelvin circulation theorem.  To
find it we need to find a generalized vorticity $\Omega(\x)$ that
commutes with the density and the metric, and at the same time transforms like a scalar
density:
\begin{subequations}\label{O-PB}
\begin{align}
  \{ n(\x), \, \Omega(\y) \} &= 0, \label{On-PB}\\
  \{ \pi_i(\x), \, \Omega(\y) \} &= \Omega(\x) \d_i\delta(\x-\y),
    \label{Opi-PB}\\
  \{G_{ij}(\x), \, \Omega(\y) \} &= 0. \label{OG-PB}
\end{align}
\end{subequations}
To satisfy Eqs.~(\ref{On-PB}) and (\ref{Opi-PB}), one can add to
$\omega$ a scalar density constructed from the metric.  This scalar
density must be chosen so that the resulting $\Omega$ has zero Poisson
bracket with the metric.  Since $\{\omega, G\}$ contains two spatial
derivatives and $\{G,G\}$ has no derivative, this scalar density has
to contain two derivatives.  The only scalar density that one can
construct from $G_{ij}$ and two derivatives is $\sqrt G\, R[G]$ where
$R[G]$ is the scalar Gaussian curvature constructed from the metric
$G_{ij}$.  By a direct check one can verify that the combination
\begin{equation}
  \Omega = \omega + \frac s2 \sqrt{G}\, R[G]
\end{equation}
satisfies Eq.~(\ref{OG-PB}).  Calculation is facilitated by working in
the coordinate systems where $G_{ij}=\delta_{ij}+\delta G_{ij}$ where
$\delta G_{ij}$ is small in the vicinity of the point under
consideration.  Hence $\Omega$ is advected by the flow:
\begin{equation}\label{Omega_conservation}
\d_t\Omega+\d_i(\Omega v^i)=0.
\end{equation}

In a later part of the paper, we will consider the application of
chiral metric hydrodynamics to the fractional quantum Hall states, in
which the fluid is coupled to a U(1) gauge field $a_\mu$.  To find a
generalization of the Kelvin circulation theorem in that case, one
notes that the equation of momentum conservation is modified by
%adding
the Lorentz force term:
%to the right-hand side:
\begin{equation}\label{NS-eb}
  \d_t \pi_i = n(e_i+\epsilon_{ij} v^j b) - n\d_i\mu -\pi_j \d_i v^j
  -\d_j(v^j\pi_i) - \d_j(\sigma^{jk}G_{ki})
  + \frac12 \sigma^{jk}\d_i G_{jk} \,.
\end{equation}
where $e_i=\d_ia_0-\d_0a_i$ and $b=\epsilon^{ij}\d_ia_j$.
One can easily check that if one redefines the generalized vorticity
as
\begin{equation}\label{Omega}
  \Omega = b + \omega + \frac s2 \sqrt{G} R[G],
\end{equation}
then it will continue to satisfy the conservation
law~(\ref{Omega_conservation}).  Hence the
quantities~(\ref{Casimirs}) remain conserved.

Another, more formal way is to consider the enlarged Poisson algebra
with the gauge potential $a_i(\x)$ and its canonical conjugate
momentum $\pi_{a_i}(\x)$ as new fields.  In the presence of a gauge
field, the Poisson bracket between $\pi_i$ is modified
\begin{equation}
  \{ \pi_i(\x), \, \pi_j(\y) \} = [\pi_j(\x)\d_i + \pi_i(\y)\d_j]\delta(\x-\y)
  - \epsilon_{ij} b(\x) n(\x)\delta(\x-\y),
\end{equation}
and two additional nontrivial Poisson brackets arise:
\begin{align}
  \{ \pi_{a_i}(\x),\, a_j(\y) \} &= \delta_{ij} \delta(\x-\y),\\
  \{ \pi_{a_i}(\x),\, \pi_j(\y) \} &= -\delta_{ij}n(\x)\delta(\x-\y).
\end{align}
One can then check that $\Omega$ defined in Eq.~(\ref{Omega})
satisfies Eqs.~(\ref{O-PB}) and has zero Poisson brackets with both
$a_i$ and $\pi_{a_i}$.

\section{Chiral metric hydrodynamics of the fractional quantum Hall fluid}

\subsection{Review of the Dirac composite fermion theory}

We now apply chiral metric hydrodynamics to describe fractional
quantum Hall fluids~\cite{Tsui:1982yy,Laughlin:1983fy}.  Let us recall
the simplest formulation of the problem: a system of interacting
nonrelativistic electrons, moving in the $xy$ plane in a magnetic field
$B$ such that the density of electrons $\rho_{\rm e}$ is smaller than
the density of states on the lowest Landau level $B/2\pi$. (Here we
absorb the factor $-|e|/c$ into the magnetic field. In the normal
convention the magnetic field directed is opposite to the $z$-axis, after
absorbing this factor it is directed along the $z$-axis.  We use the
unit system where $\hbar=1$ and define the magnetic length $\ell_{\rm
  B}=1/\sqrt B$.)

The most nontrivial feature of the quantum Hall fluids is that the
low-energy quasiparticle is completely different from the electron.
According to the composite fermion
picture~\cite{Jain:1989tx,Fradkin:1991wy,Ovchinnikov:1991,Halperin:1992mh},
near half filling, where the filling factor $\nu=2\pi\rho_{\rm e}/B\approx\frac12$, the
quasiparticle is the composite fermion, which interacts with a
dynamical gauge field $a_\mu$.  The dynamic magnetic field has
expectation value
\begin{equation}
  \< b\> = B - 4\pi \rho_{\rm e}\,.
\end{equation}
At $\nu=\frac12$ the average dynamical magnetic field is zero and the
composite fermions form a Fermi liquid. 

In the lowest Landau level limit ($B\to\infty$ at fixed $\nu$, or
purely theoretically, when the electron mass goes to zero),
particle-hole symmetry becomes an exact symmetry.  In this case the
composite fermion is a massless Dirac fermion with Berry phase of
$\pi$ around the Fermi line~\cite{Son:2015xqa}.  The composite fermion
has density $n=B/4\pi$.  This is close to, but in general not equal to the
density of the electron $\rho_{\rm e}$.
%For filling factors
%$\nu=N/(2N+1)$, $\<b\>=B/(2N+1)$ and the composite fermions form an integer
%quantum Hall state.
The composite fermion is electrically neutral but carries a nonzero
electric dipole moment $\mathbf{d}$ perpendicular  and
proportional to its momentum: $\mathbf{d}=-\ell_{\rm
  B}^2\mathbf{p}\times\mathbf{\hat z}$.

We will consider the so-called Jain states, which form two series
converging to $\nu=\frac12$,
\begin{equation}
  \nu = \frac N{2N+1}\,, \qquad \nu = \frac{N+1}{2N+1}\,, \qquad
  N=1,2,3,\ldots
\end{equation}
The $N=1$ case is the Laughlin $\nu=\frac13$ state and its
particle-hole conjugate $\nu=\frac23$, but we will be mostly
interested in the limit of large $N$, $N\gg1$.  On the Jain states,
the dynamical magnetic field has nonzero average,
\begin{equation}
  \< b \> = \pm \frac B{2N+1} \,.
\end{equation}

\subsection{Microscopic derivation of the Poisson algebra}

We now argue that at sufficiently long wavelength the quantum Hall system is
described by chiral metric hydrodynamics.  The reason is the following.
For large $N$ one can
treat the system as a Fermi liquid in a small magnetic field.  In the
bosonized approach~\cite{Haldane:1993,CastroNetoFradkin} the dynamics
of the Fermi surface can be parameterized through an infinite number of
scalar fields (called $u_n$ in Ref.~\cite{Golkar:2016thq}), one field
per spin.
Now in the regime $\omega\gg v_Fq$, where $v_F$ is the
Fermi velocity, one can truncate the tower of scalar fields to a few
first fields.  This is because at zero momentum these fields cannot
mix with each other due to the conservation of momentum, and the
strength of the mixing is governed by the parameter $v_Fq/\omega$.
This is essentially the reason the longitudinal and transverse zero
sounds in Fermi liquid with large Landau parameter $F_1$ (and
$F_\ell\sim1$ with $\ell\ge2$) can be described by a theory of
elasticity~\cite{ContiVignale,Inti:2019}.

In the fractional quantum Hall case the dynamical magnetic field has a
nonzero average value $\<b\>=\pm B/(2N+1)$, in which the composite
fermions form a integer quantum Hall state with a gap of order $\Delta
\sim b/m_*=v_F b/p_F$.  In our gapped phase $\omega\sim\Delta$ and the
mixing is small when $q\ell_B\ll 1/N$.  In this regime one can limit
ourselves fields of lowest spins.  In practice, this means spins less
or equal to 2.

\begin{figure}
  \centering
  \includegraphics[width=20em]{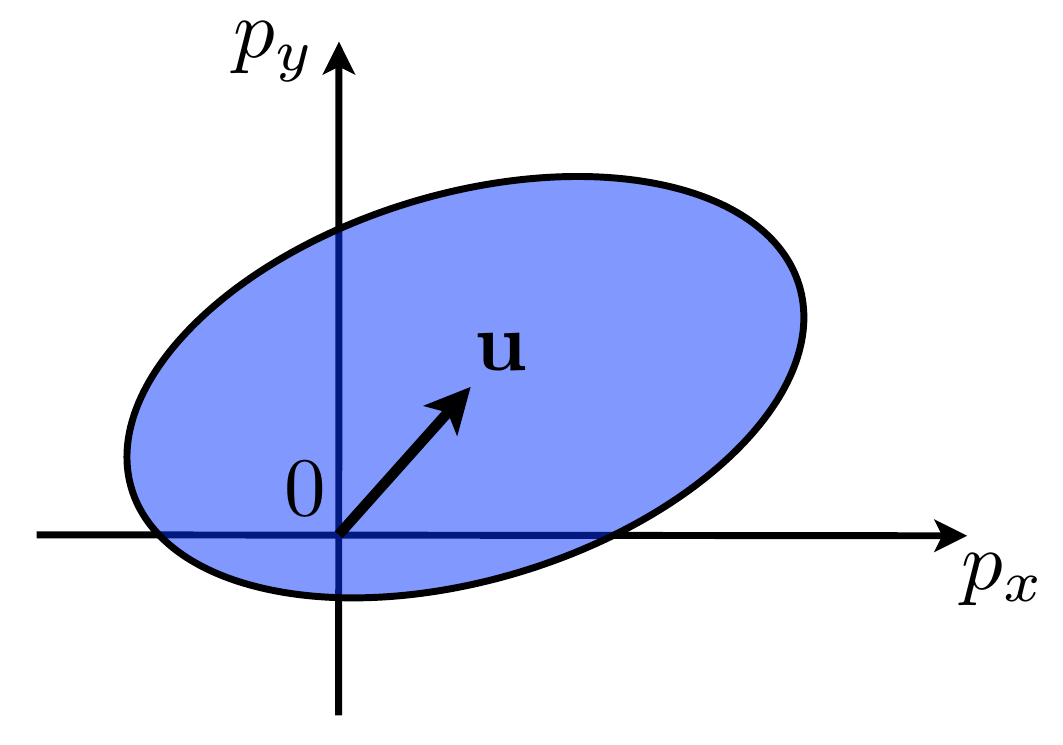}
  \caption{An elliptical Fermi surface.}
  \label{fig:fermi-surface}
\end{figure}

We assume then that the Fermi surface has the shape of an
ellipse~(Fig.~\ref{fig:fermi-surface}).  One can completely
characterize the ellipse by the moments of the distribution function
$f(\x,\p)$, which is 1 inside the ellipse and 0 outside.
%\begin{equation}
%   G^{ij} (p_i-u_i)(p_j - u_j) = 4\pi
%\end{equation}
\begin{align}
  n(\x) &= \int\!\frac{d\p}{(2\pi)^2}\, f(\x,\p), \\
  \pi_i(\x) &= \int\!\frac{d\p}{(2\pi)^2}\, p_i f(\x,\p), \\
  G_{ij}(\x) &= \frac1{\pi n(\x)}\left( \int\!\frac{d\p}{(2\pi)^2}\,
   p_i p_j f(\x,\p) - \frac{\pi_i(\x)\pi_j(\x)}{n(\x)} \right).
\end{align}
The tensor $G_{ij}$ is defined in such a way that $\sqrt{\det G}=n$.
The Poisson bracket between the fields can then be evaluate from the
following semiclassical prescription: for any two functions $A(\x,\p)$
and $B(\x,\p)$ on the phase phase,
\begin{equation}\label{PB-prescription1}
  \left\{ \int\!\frac{d\x\,d\p}{(2\pi)^d}\, A(\x,\p) f(\x,\p), \,
  \int\!\frac{d\x\,d\p}{(2\pi)^d}\, B(\x,\p) f(\x,\p) \right\}
  = \int\!\frac{d\x\,d\p}{(2\pi)^d}\, \{A, B\}(\x,\p) f(\x,\p),
\end{equation}
where $\{A,\, B\}$ is the classical Poisson bracket~\cite{LL1}
\begin{equation}\label{PB-prescription2}
  \{ A, \, B\} = \frac{\d A}{\d p_i} \frac{\d B}{\d x^i}
  -  \frac{\d A}{\d x^i} \frac{\d B}{\d p_i}
  - \epsilon_{ij} b \frac{\d A}{\d p_i} \frac{\d B}{\d p_j} \,.
\end{equation}
Following this prescription, we find that all the Poisson brackets,
except $\{G,\,G\}$, can be expressed in terms of the hydrodynamic
variables, and coincide with the expressions postulated in the
hydrodynamic theory.
%with the sole modification that the Poisson
%bracket between components of the momentum density contains a term
%proportional to the magnetic field,
On the other hand the Poisson bracket $\{G,\, G\}$ contains the third
moment of the distribution function $\int_p p_ip_j p_k f$; assuming
the Fermi surface has the shape of an ellipse, this third moment can
be expressed in terms of the lower moments,
\begin{equation}
  \int\!\frac{d\p}{(2\pi)^2}\, p_i p_j p_k f(\x,\p) =
  \frac{\pi_i \pi_j \pi_k}{n^2} +
   \pi( \pi_i G_{jk} + \pi_j G_{ik} + \pi_k G_{ij} ).
\end{equation}
At the end of a lengthy, but straightforward, calculation, one finds
\begin{equation}
  \{G_{ij}(\x), \, G_{kl}(\y)\} = - \frac {b+\omega}{\pi n}
  (\epsilon_{ik}G_{jl} + \epsilon_{il}G_{jk}
   + \epsilon_{jk}G_{il} + \epsilon_{jk} G_{ik})
  \delta(\x-\y).
\end{equation}
This is very similar to, but not exactly, Eq.~(\ref{GG-PB}): the constant
factor $1/s$ there is replaced by $(b+\omega)/\pi n$, which is a dynamical
field.  If one replaces this prefactor by its expectation value,
\begin{equation}\label{replacement}
  \frac{b+\omega}{\pi n} \to \frac{\<b\>}{\pi n} =
  \begin{cases}
    \displaystyle{\phantom{+}\frac4{2N+1}} \,, & \nu=\displaystyle{\frac N{2N+1}} \,, \vspace{6pt}\\
    \displaystyle{-\frac4{2N+1}} \,, & \nu = \displaystyle{\frac{N+1}{2N+1}}\,,
  \end{cases}  
\end{equation}
then one can identify 
\begin{equation}\label{spin}
  s = \pm \frac12\left( N+ \frac12 \right),
\end{equation}
where the upper and lower signs correspond to $\nu=N/(2N+1)$ and
$\nu=(N+1)/(2N+1)$, respectively.  The replacement can also be
justified by the following argument: up to terms containing second
derivatives of $G_{ij}$, the prefactor is $\Omega/n$, which in the
chiral metric hydrodynamics is conserved and determined completely by
the initial condition.  Provided that at infinite past the system is
in the ground state, $\Omega/n$ is then equal to $\<b\>/n$.  Note that
our semiclassical prescription for computing the Poisson bracket,
Eqs.~(\ref{PB-prescription1}) and (\ref{PB-prescription2}), cannot in
principle yield terms with second derivatives; one may hope a more
sophisticated treatment will reveal their existence.  In any case, for
the calculation of linear response, the
replacement~(\ref{replacement}) can always be made.

One may be concerned that the parameter $s$ in Eq~(\ref{spin}), which
has been previously identified with the orbital spin per particle, is neither
integer nor half-integer.  This is not a problem, as $s$ is the
\emph{average} orbital angular momentum per particle.  This suggests
another way to determine $s$.  Recall that in the composite fermion
picture, the $\nu=N/(2N+1)$ state has the composite fermions filling
up $N+\frac12$ Landau levels: half of the zero-energy Landau level
($n=0$) and the positive-energy levels with quantum numbers
$n=1,2,\ldots,N$.  Now the orbital spin of a particle on the $n$-th Landau
level is $n$, so the average orbital spin is
\begin{equation}\label{s-true}
  s = \frac1{N+\frac12} \left( \frac12\cdot 0 + 1 + 2 + \cdots + N\right)
  = \frac{N(N+1)}{2N+1} \,.
%  = \frac12 \left (N+\frac12 -\frac1{4N}+\cdots\right)
\end{equation}
For the $\nu=(N+1)/(2N+1)$ state, $s$ flips sign.  The value in
Eq.~(\ref{s-true}) does not coincides with that in Eq.~(\ref{spin}),
but the relative difference is of order $1/N^2$ at large $N$.  This
discrepancy is should be attributed to the imperfectness of the
semiclassical procedure~(\ref{PB-prescription1}).
We thus conjecture that (\ref{s-true}) represents the exact value
of the parameter $s$ and use it in later calculations.
Note that even
for $N=1$ the values of the $s$ in Eqs.~(\ref{spin}) and (\ref{s-true})
differ from each other only by a factor of 9/8.

From Eq.~(\ref{etaH}) and (\ref{s-true}) one can find the odd viscosity of the Hall fluid,
\begin{equation}\label{etaH-QH}
  \eta^H = \pm \frac{N(N+1)}{2N+1} \frac B{8\pi}\,.
\end{equation}
This should be compared with the expectation: $\eta^H=\frac14\rho_{\rm
  e}(\mathcal S-1)$ where $\mathcal S$ is the shift, and $\mathcal
S-1$ is the part of the shift associated with the guiding centers.
Equation~(\ref{etaH-QH}) exactly matches this expectation, given that $\mathcal S=N+2$ for $\nu=N/(2N+1)$ and $\mathcal S=-N+1$ for $\nu=(N+1)/(2N+1)$.

\subsection{Equations of motion of chiral metric hydrodynamics of fractional quantum Hall states}

We now derive the relevant formulas for the metric hydrodynamics of
the composite fermions.  Due to the dipolar interaction of the
composite fermion with the external electric field,
the Hamiltonian
of the composite fermion fluid contains a term which can be viewed as a source term for the
momentum density
\begin{equation}\label{H-sources}
  H = H_0[n, \pi_i, G_{ij}, b] + \int\!d\x \left( -a_0(\x) n(\x)
  + V^i(\x) \pi_i(\x) \right),
\end{equation}
where $V^i=\epsilon^{ij} E_j/B$ is the drift velocity created by the
external electric field.
This Hamiltonian transforms correctly
under Galilean boosts.  The lowest Landau level limit is
the limit of massless electrons, so the total momentum $\mathbf{P}$
is invariant under Galilean boosts, while the total energy should
transform as
\begin{equation}\label{H-Gal}
  H \to H - \bm{\beta}\cdot\mathbf{P},
\end{equation}
when going from a frame $K$ to a frame $K'$ moving with velocity
$\bm{\beta}$ with respect to $K$.  This is exactly what happens with
the Hamiltonian~(\ref{H-sources}): under boosts all variables remain
invariant except for the external electric field which transforms as
$\mathbf{E}\to\mathbf{E}+\bm{\beta}\times\mathbf{B}$, implying $V^i\to
V^i-\beta^i$ and~(\ref{H-Gal}).

In the Dirac composite fermion theory, the dynamical gauge field
$a_\mu$ appears in the action as Lagrange
multipliers enforcing constraints of the form
\begin{subequations}\label{nv-constraints}
\begin{align}
%  \frac{\delta S}{\delta a_0} &=0 \Longrightarrow
  n - \frac B{4\pi} &= 0, \label{n-constraint}\\
%  \frac{\delta S}{\delta a_i} &=0 \Longrightarrow
  nv^i - \epsilon^{ij}\d_j \frac{\delta H_0}{\delta b} &= 0.
\end{align}
\end{subequations}

The equations for the hydrodynamic variables coincide with those of
metric hydrodynamics in an electromagnetic field with
[Eqs.~(\ref{dtn-noparity}), (\ref{dtG-noparity}), and (\ref{NS-eb})],
but with the replacement $v^i\to\nu^i=v^i+V^i$.
\begin{align}
  &\d_t n + \d_i (n\nu^i) =0, \qquad \nu^i = v^i + \frac{\epsilon^{ij}E_j}B \,,
  \label{dndnnui}\\ 
  &\d_t \pi_i + n \d_i\mu + \pi_j \d_i \nu^j + \d_j (\nu^j\pi_i) +
  \d_j (\sigma^{jk} G_{ki}) - \frac12 \sigma^{jk}\d_i G_{jk}
  = n e_i + n\epsilon_{ij} \nu^j b,\\
  &\d_t G_{ij} + \pounds_\nu G_{ij}
  - \frac1s (\epsilon_{ik}G_{jl}+\epsilon_{jk}G_{il})\sigma^{kl} =0.
\end{align}
%Recall that $\mu=\delta H/\delta n$, $\tau^{ij}=2\delta H/\delta
%G_{ij}$, while $\delta H/\delta\pi_i =v^i=0$.

The problem of finding the response of the quantum Hall fluid to
external electromagnetic perturbation reduces to the finding the
solution to the hydrodynamic equations and the constraint equations in
the background $E$ and $B$ fields.  [Note that, due to the
  constraints~(\ref{nv-constraints}), Eq.~(\ref{dndnnui}) is satisfied
  automatically.]

After the solution has been found, the electron density should be read
from
\begin{equation}
  \rho_{\rm e} = \frac{\delta S}{\delta A_0} = \frac{B-b}{4\pi} -
  \epsilon^{ij}\d_i \left( \frac{\pi_j}B \right).
%  j^i_{\rm e} = \frac{\delta S}{\delta A_i} = 
\end{equation}
By using the constraint~(\ref{n-constraint}), this expression can
be rewritten as
\begin{equation}\label{rhoebomega}
   \rho_{\rm e} = \frac1{4\pi}(B-b-\omega),
\end{equation}
where $\omega$ is the vorticity defined in Eq.~(\ref{omega-def}).
One can also compute the electron current density from $j^i_{\rm
  e}=\delta S/\delta A_i$, but the formulas are somewhat cumbersome and will
not be presented here.

Two remarks on the electron density are in order.  First, one can easily
check that the Poisson bracket between the electron
density~(\ref{rhoebomega}) at two different points can be again
expressed in terms of the electron density:
\begin{equation}
  \{ \rho_{\rm e}(\x), \, \rho_{\rm e}(\y) \} =
  -\frac1{8\pi^2 n}\epsilon^{ij}\d_i\rho_{\rm e}(\x) \d_j \delta(\x-\y).
\end{equation}
This is the long-wavelength version of the Girvin-MacDonald-Platzman
algebra~\cite{Girvin:1986zz}.  The dipole contribution to the electron
density encodes a nontrivial aspect of the physics of
the lowest Landau level~\cite{Simon:1998,Nguyen:2017mcw}.

Second, using Eqs.~(\ref{nv-constraints}) and (\ref{Omega}), one can rewrite
the electron density as
\begin{equation}
  \rho_{\rm e} = \frac{B-\Omega}{4\pi} + \frac s{8\pi}\sqrt G\, R[G] .
\end{equation}
If we consider a problem in which a fractional quantum Hall system
is prepared in the ground state at $t=-\infty$
and then is perturbed by an external field,
then $\Omega$ remains proportional to the magnetic field $B$.  In this
case the perturbation of the electron density from
$\nu B/2\pi$ is given by the curvature of the dynamical metric,
\begin{equation}\label{rhoeR}
  \delta\rho_{\rm e} = \frac s{8\pi}\sqrt G\, R[G].
\end{equation}
Since the Poisson bracket of $\Omega$ with any field is either zero or
proportional to $\Omega$ itself, the Gaussian curvature density also
satisfies the long-wavelength Girvin-MacDonald-Platzman
algebra~\cite{Gromov:2017qeb}.  The identification of the Gaussian
curvature density with the electron density was part of Haldane's
proposal~\cite{Haldane:2009ke} and the bimetric theory~\cite{Gromov:2017qeb}.
Equation~(\ref{rhoeR}) will play an
important role in the later calculation of the static structure
factor.
%For now, we note that, to
%the linearized order, $\delta\rho_{\rm e}$ contains second derivatives
%of the metric tensor, hence the correlator of $\delta\rho_{\rm e}$ 

\subsection{Illustrative calculation of the density response}

We demonstrate the working of the procedure on an example of a model,
where the Hamiltonian has the quadratic form
\begin{equation}\label{Hmodel}
  H_0 = \int\!d\x\, \left[ \frac{\bm{\pi}^2}{2m_*n_0}
    + \frac\chi2 (\delta n)^2 + \frac{\tilde\mu}4 Q_{ij}^2\right],
\end{equation}
which, for the simplicity, is assumed to be independent
of $b$.  In this case the constraints~(\ref{nv-constraints}) become
\begin{equation}
  n= \frac B{4\pi}\,, \qquad v^i = 0,
\end{equation}
and the linearized hydrodynamic equations
\begin{align}
& \tilde\mu \d_j Q^{ij} = ne_i + n\epsilon_{ij}V^j b, \label{dtpi-QH}\\
& \d_t Q_{ij} + \d_i V_j + \d_j V_i - \delta_{ij} \d_k V_k
  - \frac{\tilde\mu}{sn} (\epsilon_{ik}Q_{kj} + \epsilon_{jk}Q_{ki}) = 0.
  \label{dtQ-QH}
\end{align}
Assuming that the external perturbation is $A_0(t,\x)=A_0e^{-i\omega
  t+iqx}$, then $V_y=-iqA_0/B$, and $V_{xy}=q^2A_0/B$.  Solving
Eq.~(\ref{dtQ-QH}) one finds
\begin{equation}
  Q_{xx} = - Q_{yy} = \frac{\omega_Q}{\omega^2-\omega_Q^2} V_{xy} \,,\qquad
  Q_{xy} = -\frac{i\omega}{\omega^2-\omega_Q^2} V_{xy}\,,
\end{equation}
and from Eq.~(\ref{dtpi-QH})
\begin{equation}
  e_y = \frac{\tilde\mu}n\d_x Q_{xy}\,,
\end{equation}
which means
\begin{equation}
  \nabla\times e = i\frac{\omega q^2}{\omega^2-\omega_Q^2} V_{xy}\,. 
  \end{equation}
Using $\dot b + \nabla\times e=0$ one finds
\begin{equation}
  b = \frac{q^2}{\omega^2-\omega_Q^2} V_{xy} \,,
\end{equation}
which translates to fluctuation in the density of electrons
\begin{equation}
  \delta\rho_e = - \frac b{4\pi} = - \frac{\tilde\mu}{B^2}
  \frac{q^4}{\omega^2-\omega_Q^2} A_0.
\end{equation}
Using the linear response theory, one then finds the time-ordered Green
function the electron density
\begin{equation}\label{Trhorho}
  \< \rho_{\rm e} \rho_{\rm e}\>_{\omega,q} \equiv
  \int\!dt\,d\x\, e^{i\omega t- i\mathbf{q}\cdot\mathbf{x}}
  \< T \rho_{\rm e}(t,\x),\, \rho_{\rm e}(0,\mathbf{0}) \> = 
  i \frac{\tilde\mu}{B^2}
  \frac{q^4}{\omega^2-\omega_Q^2+i\epsilon}\,.
\end{equation}
The Green function has poles at $\omega^2=\omega_Q^2$, which one can
identify with the long-wavelength part of the dispersion curve of the
magnetoroton.  Furthermore, the residue at the pole behaves like
$q^4$, which is the expected behavior for the density-density
correlator on the lowest Landau level.

One can make the model~(\ref{Hmodel}) more complicated by including a
dependence on $b$, as well as terms with derivatives like
$\pi_i\d_i n$ or $Q_{ij}\d_i\pi_j$.  Instead of doing that, we now turn to a physical
observable that is independent of the precise form of the
Hamiltonian.

\subsection{The static structure factor}

The static structure factor is an important quantitative
characteristics of the fractional quantum Hall state.  It appears, in
particular, in the variational treatment of the
magnetoroton~\cite{Girvin:1986zz}.  The procedure outlined above can
be used to compute, from a given Hamiltonian $H[n,\pi_i,G_{ij}]$, the
density response function $\<\delta\rho_{\rm e}\delta \rho_{\rm e}\>_{\omega,q}$.  By integrating
out this function over the frequency one can recover the equal-time
density-density correlator, which is proportional to the static
structure factor.  For example, from Eq.~(\ref{Trhorho}) one obtains
\begin{equation}\label{SSF-intomega}
  \< \delta\rho_{\rm e}\delta\rho_{\rm e} \>_q
  = \int\!\frac{d\omega}{2\pi}\, \<\delta\rho_{\rm e}\delta\rho_{\rm e}\>_{\omega,q}
  = \frac{\tilde\mu}{2B^2\omega_Q} q^4
  = \frac{ns}{4B^2}q^4.
\end{equation}
In particular, one notices that $\tilde\mu$ cancels out after substituting
Eq.~(\ref{omegaQ}).  This fact has a deeper reason, going beyond the
simple model (\ref{Hmodel}): the long-wavelength behavior of the
equal-time density-density correlator is determined by the Poisson
algebra, but not by the detail form of the Hamiltonian~\cite{Nguyen:2017mcw}.

To show that, we note that in the quantum theory, to quadratic order
the ``holomorphic'' component $Q\equiv
Q_{zz}=\frac14(Q_{xx}-2iQ_{xy}-Q_{yy})$ and the ``antiholomorphic''
component $\bar Q\equiv Q_{\bar z\bar
  z}=\frac14(Q_{xx}+2iQ_{xy}-Q_{yy})$ of the metric tensor have the
commutator
\begin{equation}\label{barQQcomm}
  [ \bar Q(\x), \, Q(\y) ] = \frac1 {sn}\delta(\x-\y),
\end{equation}
and hence can be consider as creation and annihilation operators.  If
$s>0$, then $Q$ is the creation and $\bar Q$ annihilation operator,
and the roles are reverse for $s<0$.
%For definiteness, let us consider $s>0$.
To quadratic order the Hamiltonian contains terms of the form $Q\bar
Q$, and the problem reduces itself in the quadratic level to the
problem of a harmonic oscillator.  Terms that does not contain one $Q$
and one $\bar Q$ appears in a very high order in the derivative
expansion, $(\bar\d^2 Q)^2$ or $(\d^2\bar Q)^2$ and can be neglected.
The vacuum is then annihilated by the annihilation operator.
\begin{equation}\label{Qvacuum}
\begin{split}  
  Q(\x) |0\> &= 0, \qquad \text{if $s>0$}, \\\
  \bar Q(\x) |0\> &= 0, \qquad \text{if $s<0$} .
\end{split}
\end{equation}

It is now straightforward to compute the equal-time correlator of the
electron density operator.  As shown before, the fluctuating part of
the electron density is proportional to the Gaussian curvature density
of the dynamical metric, i.e.,
\begin{equation}
  \delta\rho_{\rm e} = \frac s{8\pi} \sqrt G R =
  \frac s{2\pi} (\d^2 \bar Q + \bar\d^2 Q),
\end{equation}
which implies
\begin{equation}
  \<\delta\rho_{\rm e} \delta\rho_{\rm e}\>_q
  = \int\!d\x\, e^{-i{\bf q}\cdot{\bf x}} \< \delta\rho_{\rm e}(\x) \delta\rho_{\rm e}(\mathbf{0})\>
  = \frac{s^2q^4}{64\pi^2}\! \int\! d\x\, e^{-i\textbf{q}\cdot\textbf{x}}
  \left[ \<0| Q(\x) \bar Q({\bf 0}) + \bar Q(\x) Q({\bf 0}) |0\> \right].
\end{equation}
Now by using Eqs.~(\ref{barQQcomm}) and (\ref{Qvacuum}) we find
\begin{equation}    
  \<\delta\rho_{\rm e} \delta\rho_{\rm e}\>_q = \frac{|s|}{64\pi^2 n} q^4 .
\end{equation}
For $\nu=N/(2N+1)$ and $\nu=(N+1)/(2N+1)$ states, using the value of
$s$ from Eq.~(\ref{s-true}),
\begin{equation}\label{rhorhoq}
  \<\delta\rho_{\rm e}\delta\rho_{\rm e}\>_q =
  %\frac1{32\pi}\left(N+\frac12\right) \frac{q^4}B
  \frac{N(N+1)}{2N+1} \frac{q^4}{16\pi B} \,.
\end{equation}

By convention, the static structure factor is defined as
%$\bar s(q) =
%\<\delta\rho_{\rm e}\delta\rho_{\rm e}\>_q/\<\rho_{\rm e}\>$.  Our
%calculations thus imply
%for the $\nu=N/(2N+1)$ state,
\begin{equation}\label{sq-result}
  \bar s(q) = \frac{\<\delta\rho_{\rm e}\delta\rho_{\rm e}\>_q}{\<\rho_{\rm e}\>} =
%  \frac1{8N} \left(N+\frac12\right)^2 (q\ell_B)^4
  %  = \frac1{8N} \left( N^2 +N + \frac1{4}\right) (q\ell_B)^4
  \begin{cases}
    \frac18(N+1) (q\ell_B)^4, & \nu=\frac{N}{2N+1}\,,\\
    \frac18 N (q\ell_B)^4, & \nu=\frac{N+1}{2N+1}\,.
  \end{cases}
\end{equation}

In fractional quantum Hall physics one distinguishes unprojected and
projected static structure factor~\cite{Girvin:1986zz}.  As the theory
that we propose is supposed to describe the physics on a single Landau
level, the quantity computed in Eq.~(\ref{sq-result}) should be
identified with the projected static structure factor.  This can be
seen also from the calculation in Eq.~(\ref{SSF-intomega}): performing
the integral over $\omega$, one picks up only a pole at
$\omega=\omega_Q$ or $-\omega_Q$, which corresponds to an
intra-Landau-level excitation.  For the unprojected static structure
factor, the integral would picks up another contribution from the Kohn mode,
absent in our theory.

That $\bar s(q)$ behaves like $q^4$ is a known
fact~\cite{Girvin:1986zz}.  
In our theory, one can trace back the
$q^4$ dependence from the identification of $\delta\rho_{\rm e}$ with
the Gaussian curvature, which, to linearized order, is a sum of
second derivatives of metric components.

As our calculation is semiclassical in nature, Eq.~(\ref{sq-result})
is only reliable at large $N$.  One can estimate the accuracy of the
semiclassical approximation to be $1/N^2$, the discrepancy between Eqs.~(\ref{spin}) and (\ref{s-true}).
%where the leading term ($N^2$ inside
%the bracket) is the classical result and the next-to-leading term
%($N$) is the first semiclassical correction, ultimately related to the
%Berry phase of the composite fermion.  The next-to-next-to-leading
%contribution ($1/4$) should be considered as unreliable. (Note that
%even for $N=1$, this term is numerically small compared to the sum of
%leading two terms).  Throwing the last contribution away, one finds
%very simple results for the coefficient $\bar s_4$, defined as the
%limit $\bar s(q)/q^4$ as $q\to0$: $\bar s_4=\frac18(N+1)$ for
%$\nu=N/(N+1)$ and $\bar s_4=\frac18 N$ for $\nu=(N+1)/(2N+1)$.

Note that the coefficient $\bar s_4$, defined as the first coefficient
in the momentum expansion of the static structure factor: $\bar
s(q)=\bar s_4(q\ell_B)^4$, saturates the Haldane
bound~\cite{Haldane:2009ke} $\bar s_4\ge\frac18|\mathcal S-1|$, with
$\mathcal S$ being the shift.  In particular, for $N=1$ our value for
$\bar s_4$ exactly equals to that of the Laughlin
wave function~\cite{Girvin:1986zz}.  Note also that our result,
Eq.~(\ref{rhorhoq}), is automatically particle-hole symmetric, as our
formalism.
%Numerically, even
%for $N=1$, Eq.~(\ref{rhorhoq}) is larger than the value computed from
%the Laughlin wavefunction~\cite{Girvin:1986zz} only by a factor of
%9/8.  The good numeric agreement is probably due to the fact that the
%real expansion parameter at large $N$ is $1/(2N+1)$, and our approach
%manages to capture both the leading and the next-to-leading
%contributions.
In contrast, the static $\<\delta\rho_{\rm
  e}\delta\rho_{\rm e}\>_q$ correlator calculated from the modified
random phase approximation of the HLR theory is not particle-hole
symmetric, deviating from the result for the Laughlin wave function by
a factor of 2 to either side, depending on whether $\nu=1/3$ or
$\nu=2/3$~\cite{Nguyen:2017mcw}.

One word of precaution should be said at this point.  Here we have
calculated the static structure factor using our hydrodynamics, which
is a low-energy effective theory, valid below some energy cutoff
$\Lambda$, locating between the energy scale of the Coulomb
interaction $\Delta_C$ and the energy of the magnetoroton,
roughly $\Delta_C/N$.  Since the static structure factor is the
integral of the dynamic structure factor over the frequency, it is
possible that the effective field theory misses a contribution to the
integral from the Coulomb interaction energy scale $\Delta_C$.
Such a contribution should be independent of $N$ and modifies
Eqs.~(\ref{rhorhoq}) and (\ref{sq-result}) to
\begin{equation}\label{rhorhoq-c}
  \<\delta\rho_{\rm e}\delta\rho_{\rm e}\>_q =
  \left[\frac{N(N+1)}{2N+1}+c\right] \frac{q^4}{16\pi B} \,.
\end{equation}
and
\begin{equation}\label{sq-result-c}
  \bar s_4 = 
  \begin{cases}
    \frac18 [N+1 + c + O(N^{-1})], & \nu=\frac{N}{2N+1}\,,\\
    \frac18 [N +c + O(N^{-1})], & \nu=\frac{N+1}{2N+1}\,.
  \end{cases}
\end{equation}
where $c$ is a positive number.  From our experience with the Laughlin
case $N=1$, one can expect $c$ to be numerically small.  The large $N$
asymptotics of $s_4$, as well as difference between $s_4$ at
$\nu=N/(2N+1)$ and $\nu=(N+1)/(2N+1)$ remains model-independent
predictions of our calculation.

We conclude by noting here that the excited state for $\nu=N/(2N+1)$
is obtained by acting the holomorphic component of the metric $Q$ on
the ground state: $Q|0\>$ and has angular momentum $2$ (directed along
the $z$ direction, i.e., in the direction opposite to the magnetic
field).  For the $\nu=(N+1)/(2N+1)$ state, the magnetoroton state is
$\bar Q|0\>$ and has angular momentum $-2$.  As previously suggest,
the spin of the magnetoroton is, in principle, measurable in polarized
Raman scatterings, with the incoming and outgoing photons carrying
opposite spin of $\pm1$~\cite{Golkar:2013gqa,Liou:2019}.  One
potential issue is that resonant Raman scatterings involve the hole
bands of GaAs which has the square lattice symmetry, but not the full
rotational symmetry, making spins 2 and $-2$ indistinguishable from
the symmetry point of view.  Fortunately, detailed calculations using
Luttinger-Kohn Hamiltonian with realistic Luttinger parameters show
that the mixing between spins 2 and $-2$ in a polarized Raman
scatterings is suppressed by a large numerical
factor~\cite{Dung-Raman}.  We are unaware of any previous attempt to
measure the spin of the magnetoroton experimentally.

\section{Conclusion}

In this paper we have introduced, at first as a purely mathematical
construction, a new hydrodynamic theory called ``chiral metric
hydrodynamics,'' in which the set of hydrodynamic degrees of freedom
is extended to include a metric with components forming a
$\mathfrak{sl}(2,\mathbb{R})$ algebra under the Poisson brackets.  The
resulting medium has a characteristic frequency $\omega_Q$ at which
the metric components rotate.  At frequencies small compared to
$\omega_Q$, the medium is a fluid with odd viscosity; at frequencies
high compared to $\omega_Q$, it behaves like a solid.

We argue that this chiral metric hydrodynamics, coupled to a
dynamical gauge field, provides a convenient and reliable description
of the long-distance dynamics of the quantum Hall Jain states near
half filling.  Using the Kelvin circulation theorem, we establish the
relationship between the electron density and the Gaussian curvature
density.  This relationship allows one to derive the coefficient of the
leading $q^4$ asymptotics of the static structure factor at low
momenta.

It would be interesting to understand whether the hydrodynamic
framework can be extended to correctly reproduce the next $q^2$
corrections to correlation functions.
%Since the
%Girvin-MacDonald-Platzman does not have correction at the $q^2$ level,
%one may expect that the relationship between the electron density and
%the momentum density does not need modification.
It has been argued that, for a class of quantum Hall states, the $q^6$
term in the static structure factor has topological
nature~\cite{Can:2014ota,Can:2014awa,Nguyen:2016hqh}.
In Ref.~\cite{Nguyen:2017qck} it has been argued that these corrections
are related to spin-3 modes.
The Jain states
around $\nu=1/4$ and $\nu=3/4$ are also interesting subjects of study
using chiral metric hydrodynamics.  As there is no particle-hole
symmetry (see, however, Ref.~\cite{Goldman:2018bxb}) the low-energy
effective Lagrangian generically contains a Chern-Simons term $ada$,
making the constraint equations slightly more complicated.  It seems
that one can relate the coefficient $\bar s_4$, more precisely, the
difference between $\bar s_4$ on two Jain states $\nu=N/(4N+1)$ and
$\nu=N/(4N-1)$ at large $N$ with the Berry phase of the composite
fermion~\cite{Dung-Son-unpublished}.

It is also interesting to understand if the chiral metric hydrodynamic
theory proposed here can be applied to other physical systems.  A
candidate is a $p_x+ip_y$ superfluid near a nematic phase transition.
A particularly interesting possibility is that the theory is applicable
to the Pfaffian and anti-Pfaffian states.  A measurement of the
spectrum of neutral excitations in the bulk, in particular the spin of
the magnetoroton, would help the determination of the nature of the
$\nu=\frac52$ state~\cite{Banerjee:2017}. In is also interesting to
investigate supersymmetric extensions of the chiral metric hydrodynamics
theory, which may provide an interpretation of the fermionic
counterpart of the magnetroton~\cite{Bonderson:2011,Moeller:2011}.

Finally, it would be very interesting to know if chiral metric
hydrodynamics can be realized in soft-matter systems, e.g., through
designing an appropriate active fluid~\cite{Vitelli:2017}. This will
bring about an interesting connection between fractional quantum Hall
and classical soft-matter systems.

\acknowledgments

The author thanks Alexander Abanov, Michel Fruchart, Paolo Glorioso,
Andrei Gromov, Matthew Lapa, Umang Mehta, Vincenzo Vitelli, and Paul
Wiegmann for discussions, and Dung Xuan Nguyen for discussion and comments
on a previous version of this manuscript.  This work is supported, in part, by DOE
grant No.\ DE-FG02-13ER41958, a Simons Investigator grant from the
Simons Foundation, and a Big Ideas Generator (BIG) grant from the
University of Chicago.

%\appendix

%\section{Landau Fermi liquid theory as a higher-spin theory}

\end{document}